\documentclass[11pt,letterpaper]{article}
\usepackage[utf8]{inputenc}
\usepackage{preamble}

\title{Photon Statistics from Yb$^{3+}$-Doped CsPbCl$_3$ are Inconsistent with Quantum Cutting}

\author{
Vincent R.M. Benning,
Faris Horani,
Daniel R. Gamelin,
Freddy T. Rabouw$^\ast$
}

\date{}

\begin{document}

\maketitle

\begin{abstract}
CsPb(Cl$_{1-x}$Br$_x$)$_3$:Yb$^{3+}$ has been widely reported as a broadband quantum-cutting material with a photoluminescence quantum yield exceeding 100\%, making it a promising candidate for enhancing the blue--green spectral response of silicon photovoltaics. Many groups have reproduced absolute photoluminescence quantum yields over 100\%, but others have struggled to obtain such high values. Here, we test the quantum-cutting capabilities of CsPbCl$_3$:Yb$^{3+}$ nanocrystals and bulk material using photon-correlation analysis. A quantum-cutting material is expected to exhibit photon bunching, but our experiments on CsPbCl$_3$:Yb$^{3+}$ show no such behavior. In fact, we observe the opposite---anti-bunching---under focused-excitation conditions. This observation can be explained with the previously established Auger-quenching pathway in CsPbCl$_3$:Yb$^{3+}$. Our results thus confirm high-power Auger quenching but question earlier descriptions of quantum cutting in CsPb(Cl$_{1-x}$Br$_x$)$_3$:Yb$^{3+}$.
\end{abstract}

\clearpage

Quantum-cutting phosphors are spectral converters that transform a high-energy photon absorbed into two or more lower-energy photons emitted. The energy efficiency can approach unity, thus avoiding conventional Stokes losses. When integrated with a photovoltaic (PV) cell, such materials can increase the photocurrent by converting high-energy solar photons into multiple lower-energy photons before absorption by the cell.\cite{congreve2013external, van2009lanthanide, semonin2011peak,shockley2018detailed}

To this end, Yb$^{3+}$ doped CsPbCl$_3$ perovskites have attracted particular interest as a promising quantum cutter. These materials combine the strong and broadband absorption of the semiconductor host with the narrow-line emission in the near-infrared (NIR) of Yb$^{3+}$.\cite{zhou2017cerium,pan2017doping,cohen2019quantum,milstein2018picosecond,crane2019detailed,zhang2018yb} Moreover, by partial anion exchange of Cl$^-$ to Br$^-$ the band gap can be tuned to exactly two times the Yb$^{3+}$ excited-state energy\cite{nedelcu2015fast,guhrenz2016solid}. These properties would make the materials ideal for integration with mature silicon-based PV technology. 

The quantum-cutting capability of CsPbCl$_3$:Yb$^{3+}$ was identified from integrating-sphere measurements revealing photoluminescence quantum yields (PLQYs) exceeding 100\%.\cite{zhou2017cerium,pan2017doping,milstein2018picosecond,kroupa2018quantum, luo2018quantum,zhou2019impact} 
Additionally, spectroscopic data are consistent with a quantum-cutting mechanism involving concerted nonradiative excitation of two Yb$^{3+}$ ions following photoexcitation of the perovskite host, most notably, the strong dependence of the intensity of NIR Yb$^{3+}$ emission on the spectral overlap of the perovskite emission and the absorption associated with double excitation of a pair of Yb$^{3+}$ ions.\cite{roh2023negative,milstein2019anion}
Furthermore, the observation of a sub-bandgap trap state in CsPbCl$_3$ doped with trivalent lanthanides provided a plausible intermediate state for the energy-transfer process from the host exciton to two Yb$^{3+}$ ions.\cite{milstein2018picosecond, roh2020yb, sommer2022defect, li2019mechanism, tepliakov2024trap}

At the same time, attempts to reproduce absolute PLQYs approaching 200\%  have yielded inconsistent results, with reported values frequently below unity.\cite{van2025cspbcl3, roh2020yb,zhao2020room,ishii2020sensitized,tepliakov2024trap,ye2023980,cleveland2023physical, stefanski2021optical,demkiv2024effect, rubio2026defect} Supplementary Information Table S1 summarizes the PLQY values reported by various groups. While PLQYs below 100\% from integrating-sphere experiments raise questions about the quantum-cutting capabilities of CsPbCl$_3$:Yb$^{3+}$, they could also be due to defects in the particular material batch causing nonradiative losses.\cite{timmerman2011step,klimov2014multicarrier} To resolve this ambiguity, a direct experimental probe for quantum cutting is required.

Here, we employ photon-correlation analysis as an unambiguous test of quantum cutting in Yb$^{3+}$-doped CsPbCl$_3$ nanocrystal (NC) ensembles and bulk powders.\cite{benning2025photon,de2017non} 
In a quantum-cutting material, absorption of a single high-energy photon creates two excited states that will emit in short succession. Under continuous-wave illumination this results in an enhanced probability of detecting pairs of photons on the timescale of the emitter excited-state lifetime. However, our photon-correlation analysis reveal no such photon bunching. This observation challenges previous descriptions of quantum cutting in Yb$^{3+}$-doped metal halide perovskites. Under focused excitation, we observe the opposite of photon bunching: photon anti-bunching. This behavior is consistent with previously studied Auger quenching of the host exciton by excited Yb$^{3+}$ ions.\cite{roh2023evolution, roh2020yb, erickson2019photoluminescence} This work underpins the importance of photon-statistics-based measurements to interrogate quantum cutting capabilities, and highlights that photon statistics can reveal other non-linear excited-state processes.
\\

\begin{figure}[!t]
    \centering
    \includegraphics[width=\linewidth]{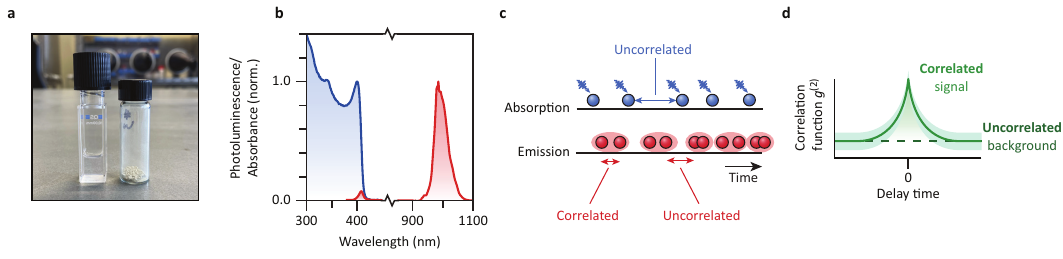}
    \caption{\textbf{CsPbCl$_3$:Yb$^{3+}$ samples and photon-correlation concept. (a)} Photograph of CsPbCl$_3$:Yb$^{3+}$ synthesized via colloidal methods (nominally 12\% Yb$^{3+}$) dispersed in hexane (left) and mechanochemical methods (nominally 4\% Yb$^{3+}$) as a microcrystalline powder (right). \textbf{(b)} Representative absorbance spectrum (blue) and photoluminescence spectrum under excitation at $\lambda_\text{exc} =$ 375~nm (red). \textbf{(c)} Illustration of hypothetical photon stream from a quantum-cutting material. Continuous-wave excitation produces temporally uncorrelated absorption events. Photon pairs originating from the same absorption event (light red ellipse) are temporally correlated, while photons from different absorption events are not. \textbf{(d)} Schematic illustration of the expected second-order correlation function showing photon bunching. Uncorrelated photons contribute to the flat background, whereas correlated photon pairs give rise to a super-Poissonian peak at zero delay time. The noise on the correlation function is indicated by the teal band.}
    \label{fig:Main1}
\end{figure} 

We study two types of CsPbCl$_3$:Yb$^{3+}$ samples: one NC sample synthesized via a colloidal method and one microcrystalline powder synthesized via a mechanochemical method (Figure \ref{fig:Main1}a). The nominal Yb$^{3+}$ doping concentrations are 12\% for the colloidal sample and 4\% for the mechanochemically synthesized sample. SI sections S1.1--1.4 show the synthesis procedures and characterization of the samples. Upon excitation of the CsPbCl$_3$ host at 375~nm the samples show moderate excitonic emission in the visible around 410~nm and strong NIR emission centered around 980~nm  corresponding to the Yb$^{3+}$ $^2$F$_{5/2}\rightarrow^2$F$_{7/2}$ transition (Figure \ref{fig:Main1}b). 

\begin{figure}[!t]
    \centering
    \includegraphics[width=\linewidth]{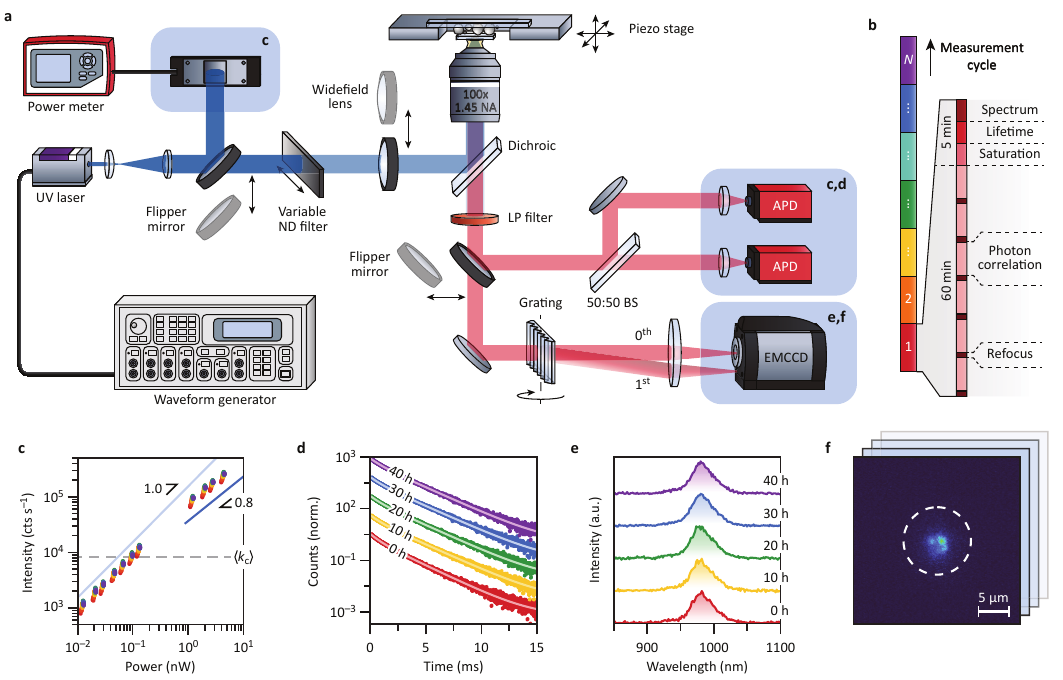}
    \caption{\textbf{Automated data-acquisition setup. (a)} Schematic overview of the experimental setup. All optical elements indicated by arrows are motorized and can be rotated or moved into/out of the optical path by the automated measurement routine. Detectors used for the data in panels (c–f) are labeled accordingly. (c) Excitation intensity is measured in the excitation path using a power meter. (c,d) Avalanche photodiodes (APDs) collect Yb$^{3+}$ emission with nanosecond temporal resolution for lifetime and correlation measurements. (e,f) The electron-multiplying CCD camera collects an image of the Yb$^{3+}$ emission for autofocus, or its spectrum if a grating is moved in place. \textbf{(b)} Timeline of the automated measurement routine. \textbf{(c)} Repeated saturation curves (dots), colors from red to purple corresponds to data recorded over 40 h of measurement duration. Solid lines are guides to the eye with power exponents of 1 and 0.8. The dashed line indicates the average count rate used during the photon-bunching experiment. \textbf{(d)} photoluminescence decay curves (dots) of Yb$^{3+}$ emission ($\lambda_\text{em} >$ 900~nm) under excitation at $\lambda_\text{exc} =$ 375~nm. \textbf{(e)} Emission spectra of Yb$^{3+}$ at different measurement times. Patterns are offset for clarity. \textbf{(f)} Image of Yb$^{3+}$ photoluminescence of an aggregate of CsPbCl$_3$:Yb$^{3+}$ NCs. The white circle shows the projection of the APD pixel onto the sample plane.}
    \label{fig:Main2}
\end{figure}

To test whether the observed NIR emission stems from quantum cutting, we employ photon-correlation analysis.\cite{benning2025photon,de2017non} In an ideal quantum-cutting material, absorption of a single high-energy photon results in the emission of two photons. In CsPbCl$_3$:Yb$^{3+}$, spectroscopic studies have suggested a mechanism involving concerted excitation of two Yb$^{3+}$ ions.\cite{roh2023negative,milstein2019anion} In this scenario, under continuous-wave excitation of an ensemble of emitters, two photons generated following the same absorption event are emitted in short succession and are therefore temporally correlated. This should give rise to a bunching peak centered around zero delay in the second-order correlation function ($g^{(2)}$) (Fig.~\ref{fig:Main1}c,d). In contrast, emission events originating from different absorption events are uncorrelated in time, resulting in a (nearly) flat background in the $g^{(2)}$. The visibility of the bunching feature is limited by shot noise on the background. In practice, collection losses in the detection setup and nonradiative losses in the sample substantially reduce the detected fraction of correlated photon pairs. Additionally, if one photon of a pair is lost, the remaining photon contributes to the uncorrelated background. Consequently, the background level increases while simultaneously the amplitude of the bunching feature decreases. We carefully optimize and characterize our detection setup (SI section S2). Nevertheless, acquisition times of several tens of hours are required to accumulate sufficient statistics to reliably determine the presence or absence of a bunching feature.

\begin{figure}[!t]
    \centering
    \includegraphics[width=\linewidth]{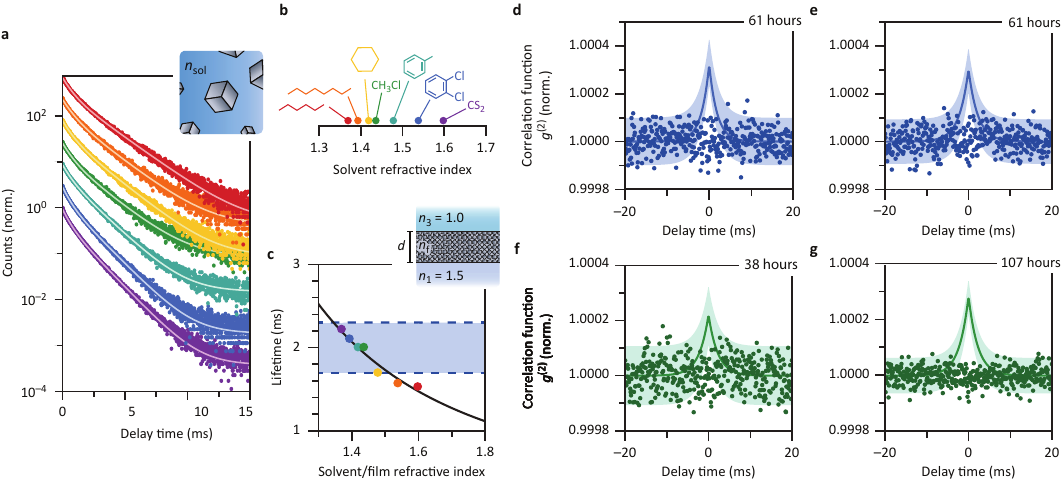}
    \caption{\textbf{Dynamics and photon statistics of the Yb$^{3+}$ emission. (a)} photoluminescence decay curves (dots) of CsPbCl$_3$:Yb$^{3+}$ NCs dispersed in solvents of varying refractive index upon excitation at $\lambda_\text{exc} =$ 375~nm and recorded at $\lambda_\text{em} >$ 900~nm. The solid lines are biexponential fits. Curves are offset for clarity. \textbf{(b)} Refractive indices of solvents used for NC dispersions. \textbf{(c)} Slow lifetime component from the fits in (a) (dots). The black line shows the fit to eq~\ref{eq: kyb total}. The dark blue dashed lines show the shortest and longest experimental radiative lifetimes measured for the different samples on a glass substrate. \textbf{(d–g)} Experimental normalized second-order correlation function (dots) for \textbf{(d,e)} two colloidally synthesized samples and \textbf{(f,g)} two mechanochemically synthesized samples. Solid lines show the expected correlation function assuming quantum cutting, based on the emission efficiencies and radiative lifetimes from photoluminescence decay experiments. The band indicates the $\pm2\sigma$ noise on the expected bunching signal. No experimental bunching signal is observed.}
    \label{fig:Main3}
\end{figure}

We perform photon-correlation measurements using the automated optical setup shown in Fig.~\ref{fig:Main2}a. An ultraviolet continuous-wave laser ($\lambda_\text{exc} = 375$~nm) illuminates the sample via a microscope objective, either in wide field or focused. The same objective collects the Yb$^{3+}$ photoluminescence, which is spectrally filtered to reject reflected excitation light and exciton emission using a dichroic mirror and long-pass filter. Subsequently the light is sent to two single-photon avalanche photodetectors (APDs). The $g^{(2)}$ curve is then obtained from the cross-correlation of the photon streams by the two detectors (Fig.~\ref{fig:Main3}d–g). 

An automated measurement routine ensures that the sample remains in focus
during the multi-hour acquisition period and checks sample integrity at regular intervals. To probe the integrity we record hourly photoluminescence saturation curves, time-resolved photoluminescence decay curves, emission spectra, and NIR photoluminescence images (Fig.~\ref{fig:Main2}b--e). Further details of the optical setup and measurement routine, as well as the full extended dataset for all samples, are provided in SI sections S1.5--1.6 and S2--3. 

The power-dependent measurements (Fig.~\ref{fig:Main2}c) confirm stable Yb$^{3+}$ emission at the count rates employed in the photon-bunching experiments, well below the onset of significant Auger quenching. Time-resolved Yb$^{3+}$ decay curves recorded at $\lambda_\text{em} > 900$ nm (Fig.~\ref{fig:Main2}d) are well described by a biexponential function, with both lifetime components remaining constant throughout the experiment, indicating sample stability. In addition, NIR photoluminescence images and spectra (Fig.~\ref{fig:Main2}e–f) show no systematic decrease in emission intensity over time, nor any discernible changes in spectral shape or peak position during the measurement, further demonstrating sample stability.
\\

Before we present and analyze the experimental photon correlation, we will calculate the expected bunching magnitude under the assumption that CsPbCl$_3$:Yb$^{3+}$ is indeed a quantum cutter. In the absence of significant Auger quenching (SI section S4.7) the expected bunching at $\tau = 0$ is given by

\begin{equation}
g^2(0)-1 = \frac{k_\text{r} \eta_\text{col}}{2 I^{(\infty)}},
\end{equation} 
where $k_\text{r}$ is the radiative decay rate of the Yb$^{3+}$ excited state, $\eta_\text{col}$ is the collection efficiency of our setup (SI section S2) and $I^{(\infty)}$ is the steady-state photon count rate (SI section S3). 

The bunching amplitude depends directly on $k_\text{r}$. The value of the radiative rate for Yb$^{3+}$ in bulk CsPbCl$_3$, $k_\text{r}^\text{bulk}$, is not exactly known. Moreover, for Yb$^{3+}$ in an inhomogeneous optical environment, $k_\text{r}$ may be different from a homogeneous large bulk crystal. This is quantified through the local density of optical states (LDOS), which determines deviations of $k_\text{r}$ from the homogeneous bulk value $k_\text{r}^\text{bulk}$.  We will make use of the known scaling of the LDOS with solvent refractive index to determine $k_\text{r}^\text{bulk}$. Next, we will use $k_\text{r}^\text{bulk}$ to estimate $k_\text{r}$ for a film of Yb$^{3+}$-doped NCs or microcrystals lying on a glass substrate.

Specifically, we measure the Yb$^{3+}$ decay dynamics in NC dispersions prepared with solvents of varying refractive indices, $n_\text{sol}=1.38$–$1.59$ (Figure~\ref{fig:Main3}a,b). For NCs in a dispersion, the scaling of the LDOS with refractive index---and therefore the scaling of $k_\text{r}$ with refractive index---is known.\cite{senden2015photonic,wang2017photonic} The Yb$^{3+}$ transition has both magnetic-dipole and electric-dipole character and in general, nonradiative decay is possible.\cite{dodson2012magnetic,rabouw2016europium} The total decay rate scales as

\begin{equation}
k_\text{Yb}(n_\text{sol}) = 
\frac{n_\text{sol}^3}{n_\text{NC}^3}  k_\text{MD}  +
\frac{n_\text{sol}}{n_\text{NC}}\left(  \frac{3n_\text{sol}^2}{2n_\text{sol}^2+n_\text{NC}^2} \right)^2
k_\text{ED}  + k_\text{nr}
\label{eq: kyb total}
\end{equation}

where $k_\text{ED}$ and $k_\text{MD}$ are the magnetic- and electric-dipole radiative decay rates in bulk CsPbCl$_3$, respectively, and $k_\text{nr}$ is the nonradiative decay rate, which we assume to be solvent-independent.

Figure~\ref{fig:Main3}c shows the experimental slow lifetime component as a function of solvent refractive index, revealing a clear dependence consistent with LDOS-induced modification of the radiative decay rate. Fitting Equation \ref{eq: kyb total} to this, while keeping $k_\text{MD}$ fixed to the tabulated value of $0.087$~ms$^{-1}$,\cite{dodson2012magnetic} yields $k_\text{ED} = 0.71$~ms$^{-1}$ and $k_\text{nr} = 0.025$~ms$^{-1}$.

Using the fitted value of $k_\text{ED}$ and tabulated value of $k_\text{MD}$  we can estimate $k_\text{r}$ for NCs on a glass substrate with air above (Figure~\ref{fig:Main3}c). In this three-layer geometry, the LDOS depends on the effective refractive index $n_\text{eff}$ of the NC layer (constituting NCs, ligands, and air gaps), on the local thickness $d$ of the NC layer, and on the location of an Yb$^{3+}$ within the NC layer.\cite{karaveli2011spectral} Taking reasonable estimates of $n_\text{eff} = 1.59$ and averaging over $d = 250$–$4000$~$\mathrm{\mu}$m and over emitters positioned at all locations within the layer, we estimate that $k_\text{r}$ = 2.3 ms$^{-1}$. This value is used for $k_\text{r}$ in calculations of the expected bunching amplitude throughout the remainder of this work. \cite{zhang2024optimizing} SI section S1.7 explains the three-layer calculation in more detail. Due to local variations within the sample composition i.e. layer thickness and distribution of emitters within the slab as well as position-dependent local-field effect in non-spherical NCs, the exact $k_\text{r}$ will vary between NCs, which explains the multiexponential decay observed in Figure \ref{fig:Main2}d and \ref{fig:Main3}a.\cite{mangnus2021finite}
\\

Figure \ref{fig:Main3}d–g show the theoretical prediction of the $g^{(2)}$ for four different samples under the assumption of quantum cutting with the shading showing the expected $\pm2\sigma$ noise level. The experimentally measured normalized second-order correlation functions for the NC films and microcrystalline samples are overlaid in the  panels. In contrast to the predictions, none of the measurements show a bunching feature above the background noise level. Hence, the photon statistics of CsPbCl$_3$:Yb$^{3+}$ emission are inconsistent with quantum cutting according to previously proposed mechanisms.

To place this null result in context, we note that the same experimental setup and analysis procedure have previously resolved photon bunching in the established quantum-cutting material YPO$_4$:Tb$^{3+}$,Yb$^{3+}$.\cite{vergeer2005quantum,benning2025photon} This confirms that the setup works as expected and the absence of photon bunching from CsPbCl$_3$:Yb$^{3+}$ is a property of the material.

Can the absence of photon bunching be reconciled with any previously unconsidered quantum-cutting mechanism in CsPbCl$_3$:Yb$^{3+}$? Any pathway that yields two Yb$^{3+}$ emissions within a few ms would produce photon bunching, irrespective of whether the pathway is coherent or incoherent\cite{mukherjee2026mechanistic} or whether it is concerted or stepwise transfer.\cite{pan2017doping,zhang2018yb} Only a quantum-cutting mechanism that stores part of the excitation energy for times much longer than a few ms could in principle explain our current data. The integrated correlation signal would spread over a wide range of delay times and disappear into the noise background. Such explanation would however produce a slow component with significant amplitude in the photoluminescence decay experiments (Figures~\ref{fig:Main2}d,\ref{fig:Main3}a). One might also expect different photoluminescence decay curves for host excitation and resonant Yb$^{3+}$ excitation, which we have not yet observed. At present, we cannot think of a quantum-cutting mechanism consistent with both our photon-correlation results and available photoluminescence decay data.
\\

\begin{figure}[!t]
    \centering
    \includegraphics[width=\linewidth]{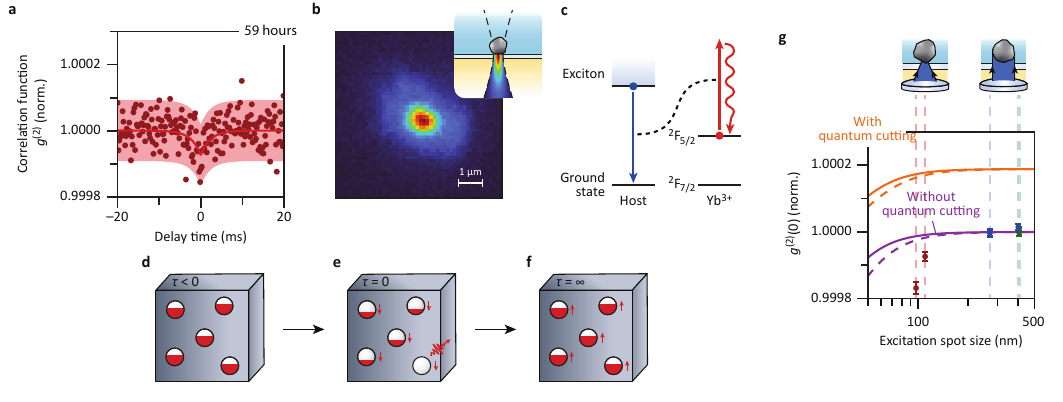}
    \caption{\textbf{Antibunching under focused excitation. (a)} Normalized second-order correlation function measured on a CsPbCl$_3$:Yb$^{3+}$ microcrystal under focused laser excitation at $\lambda_\text{exc} =$ 405~nm. The solid line is a fit to an exponentially decaying anti-bunching peak with amplitude $g^{(2)} - 1 = (-7\pm 1)\times 10^{-4}$. \textbf{(b)} Image of Yb$^{3+}$ photoluminescence of the focused emission spot. Inset shows a schematic of the laser focused on a microcrystal. \textbf{(c)} The Auger quenching process in CsPbCl$_3$:Yb$^{3+}$: an exciton in the semiconductor host is transferred to an already excited Yb$^{3+}$, which then nonradiative relaxes back to the emitting state. The host excited-state energy is lost. \textbf{(d--f)} Schematic of a volume of CsPbCl$_3$ with several Yb$^{3+}$ dopants (circles). The fraction of red filling to the circle depicts the probability that a dopant is in its excited state, explaining the occurrence of anti-bunching due to the Auger process. The dopants have a steady-state excited-state probability long before (d) or after (f) we detect a photon. The detection of a photon signals that the excited-state probabilities are decreased (e). \textbf{(f)} Experimental photon (anti-)bunching, $g^{(2)}(0)$, as a function of excitation spot size. The photon count rate is kept constant between experiment, so experiments with a smaller spot size have a larger local excitation rate density. Green, blue, and red colors match the correlation data in Figs.~\ref{fig:Main3},\ref{fig:Main4}. The solid lines are the expected bunching amplitude as a function of spot size if CsPbCl$_3$:Yb$^{3+}$ does (orange) or does not (purple) exhibit quantum cutting. The model is explained in detail in SI section S4.}
    \label{fig:Main4}
\end{figure}

When experiments are performed under focused illumination (Figure~\ref{fig:Main4}a) instead of widefield illumination (Figure~\ref{fig:Main3}d–g) we observe exactly the opposite of bunching: the emission is anti-bunched. When focusing the laser, we keep the luminescence count rate constant at approximately 6500 cts/s. A finite number of Yb$^{3+}$ dopants in the excitation volume could in principle produce anti-bunching. However, we exclude the explanation for our experimental data (Figure~\ref{fig:Main4}a) as it predicts a much smaller anti-bunching amplitude. We estimate that we collect emission of $n_\textrm{tot}\approx 10^7$  Yb$^{3+}$ ions (Figure~\ref{fig:Main4}b, S5--7). The expected anti-bunching for $n_\textrm{tot}$ independent emitters, $g^{(2)}(0) -1 =-1/n_\textrm{tot}$, would be far below the noise floor and therefore not observable. 

Instead, we interpret the anti-bunching of $g^{(2)}(0)-1 \approx 10^{-4}$ as a signature of Auger quenching in Yb$^{3+}$-doped CsPbCl$_3$ (Fig.~\ref{fig:Main4}c), which was previously reported.\cite{erickson2019photoluminescence} The effect of Auger quenching on photon statistics can be understood with a classical model. Conceptually, the model is illustrated in Figure \ref{fig:Main4}d–f. SI sections S4.5–4.9 explain the considerations in detail and derive equations. Briefly, one considers the probabilities that Yb$^{3+}$ dopants are in their excited state with a rate-equation description. At constant excitation, these reach a steady-state value (Fig.~\ref{fig:Main4}d). The detection of a photon at $\tau = 0$ signals information about the emitting nanocrystal: one of its dopants is now in the ground state. It also changes the excited-state probability of all other dopants. Indeed, just before $\tau = 0$, we know that one dopant was excited. This increases Auger quenching of NC excitons (Fig.~\ref{fig:Main4}) and thus reduces feeding of all other dopants in the NC. Using the rate constants estimated by Erickson et al.\cite{erickson2019photoluminescence} the result is a decrease in the excited-state probability for all other dopants at $\tau = 0$ (Fig.~\ref{fig:Main4}e). As $\tau$ increases further, the steady state is restored (Fig.~\ref{fig:Main4}f).

Figure \ref{fig:Main4}g shows the experimental (anti-)bunching $g^{(2)}(0)$ as a function of the excitation spot size. Between experiments, we keep the luminescence count rate constant. The expected curve for the scenario \emph{without} quantum cutting (purple) matches the experimental bunching. The dependence on spot size is due to Auger quenching, which is affected by the excitation rate density and by the number of Yb$^{3+}$ dopants in the focal volume (SI section S4.10 for details). An even better match could be obtained if we further optimize the value from the Auger quenching rate or if we assume that some Yb$^{3+}$ ions are optically inactive.\cite{xu2023atomic,roh2023evolution} Future studies could use photon correlation functions to unravel mechanistic details and quantify parameter values of Auger quenching and other nonlinear processes. In this work, we are satisfied with the qualitative match between experiment and model.
\\


In summary, we have used photon-correlation analysis to directly test the quantum-cutting capabilities of CsPbCl$_3$:Yb$^{3+}$ synthesized by mechanochemical and colloidal methods. The theoretically predicted bunching signals, calculated under the assumption of quantum cutting, are inconsistent with the experimentally measured correlation functions for all samples. These results are consistent with reports of $<100\%$ photoluminescence quantum yield for CsPbCl$_3$:Yb$^{3+}$ and suggest an absence of quantum-cutting behavior in this material. Further work will be required to establish how our findings relate to previous mechanistic studies showing dependence of the NIR Yb$^{3+}$ emission intensity on the bandgap energy of the CsPbCl$_{3-x}$Br$_x$ lattice.
At the same time, we observe anti-bunching of the Yb$^{3+}$ emission under focused excitation. This feature can be understood in terms of precious reported Auger quenching, and highlights the potential of photon correlation analysis to study nonlinear excited-state processes.

\section*{Associated content}
\subsection*{Supporting information}
Additional experimental details on sample fabrication and characterization,  experimental methods, detector efficiency, additional data correlation experiments.

\section*{Data availability}
The source data that supports the findings in this work can be downloaded from the Zenodo server (to be added)

\subsection*{Author Information}
\subsubsection*{Corresponding Author}
\noindent\textbf{Freddy T. Rabouw} -- Soft Condensed Matter and Biophysics, Debye Institute for Nanomaterials Science, Condensed Matter and Interfaces, Debye Institute for Nanomaterials Science, Utrecht University, Princetonplein 1, 3584CC Utrecht, the Netherlands; \orcidlinkf{0000-0002-4775-0859}; Email: F.T.Rabouw@uu.nl \\

\subsubsection*{Authors}
\noindent\textbf{Vincent R.M. Benning} -- Soft Condensed Matter and Biophysics, Debye Institute for Nanomaterials Science, Condensed Matter and Interfaces, Debye Institute for Nanomaterials Science, Utrecht University, Princetonplein 1, 3584CC Utrecht, the Netherlands; \orcidlinkf{0009-0006-2915-2114} \\

\noindent\textbf{Faris Horani} -- Department of Chemistry, University of Washington, Seattle, Washington 98195, United States; \orcidlinkf{0000-0001-8333-525X} \\

\noindent\textbf{Daniel R. Gamelin} -- Department of Chemistry, University of Washington, Seattle, Washington 98195, United States; \orcidlinkf{0000-0003-2888-9916}

\subsection*{Notes}
The authors declare no conflicting interests.

\section*{Acknowledgments}
This work was supported by NWO grant Vi.Vidi.203.031 (V.R.M.B., main applicant F.T.R.).The authors gratefully acknowledge postdoctoral fellowship support for F.H. from the Washington Research Foundation (WRF).

\bibliographystyle{naturemag}                     
\bibliography{ref}

@article{shockley2018detailed,
  title={Detailed Balance Limit of Efficiency of p--n Junction Solar Cells},
  author={Shockley, William and Queisser, Hans J},
  journal={J. Appl. Phys},
  volume={32},
  pages={510},
  year={1961}
}

@article{van2009lanthanide,
  title={Lanthanide ions as spectral converters for solar cells},
  author={Van Der Ende, Bryan M and Aarts, Linda and Meijerink, Andries},
  journal={Phys. Chem. Chem. Phys.},
  volume={11},
  number={47},
  pages={11081--11095},
  year={2009},
  publisher={Royal Society of Chemistry}
}

@article{semonin2011peak,
  title={Peak external photocurrent quantum efficiency exceeding 100\% via MEG in a quantum dot solar cell},
  author={Semonin, Octavi E and Luther, Joseph M and Choi, Sukgeun and Chen, Hsiang-Yu and Gao, Jianbo and Nozik, Arthur J and Beard, Matthew C},
  journal={Science},
  volume={334},
  number={6062},
  pages={1530--1533},
  year={2011},
  publisher={American Association for the Advancement of Science}
}

@article{congreve2013external,
  title={External quantum efficiency above 100\% in a singlet-exciton-fission--based organic photovoltaic cell},
  author={Congreve, Daniel N and Lee, Jiye and Thompson, Nicholas J and Hontz, Eric and Yost, Shane R and Reusswig, Philip D and Bahlke, Matthias E and Reineke, Sebastian and Van Voorhis, Troy and Baldo, Marc A},
  journal={Science},
  volume={340},
  number={6130},
  pages={334--337},
  year={2013},
  publisher={American Association for the Advancement of Science}
}

@article{cohen2019quantum,
  title={Quantum-cutting {Yb}\textsuperscript{3+}-doped perovskite nanocrystals for monolithic bilayer luminescent solar concentrators},
  author={Cohen, Theodore A and Milstein, Tyler J and Kroupa, Daniel M and MacKenzie, J Devin and Luscombe, Christine K and Gamelin, Daniel R},
  journal={J. Mater. Chem. A},
  volume={7},
  number={15},
  pages={9279--9288},
  year={2019},
  publisher={Royal Society of Chemistry}
}

@article{crane2019detailed,
  title = {Detailed-balance analysis of {Yb}\textsuperscript{3+}:{CsPb}({Cl}\textsubscript{1--x}{Br}\textsubscript{x})\textsubscript{3} quantum-cutting layers for high-efficiency photovoltaics under real-world conditions},
  author = {Crane, Matthew J. and Kroupa, Daniel M. and Gamelin, Daniel R.},
  journal = {Energy Environ. Sci.},
  volume = {12},
  number = {8},
  pages = {2486--2495},
  year = {2019},
  publisher = {Royal Society of Chemistry}
}

@article{timmerman2011step,
  title={Step-like enhancement of luminescence quantum yield of silicon nanocrystals},
  author={Timmerman, D and Valenta, J and Dohnalov{\'a}, K and De Boer, WDAM and Gregorkiewicz, T},
  journal={Nat. Nanotechnol.},
  volume={6},
  number={11},
  pages={710--713},
  year={2011},
  publisher={Nature Publishing Group UK London}
}

@article{klimov2014multicarrier,
  title={Multicarrier interactions in semiconductor nanocrystals in relation to the phenomena of Auger recombination and carrier multiplication},
  author={Klimov, Victor I},
  journal={Annu. Rev. Condens. Matter Phys.},
  volume={5},
  number={1},
  pages={285--316},
  year={2014},
  publisher={Annual Reviews}
}

@article{benning2025photon,
  title={Photon Statistics as a Tool to (Dis) Prove Cooperative Energy Transfer Quantum Cutting in Near-Infrared Emitting Materials},
  author={Benning, Vincent R M and van de Mortel, Nils and Waakop Reijers, Midas and Mastwijk, Maurits and Vonk, Sander J W and Meijerink, Andries and Rabouw, Freddy T},
  journal={ACS Energy Lett.},
  volume={10},
  number={9},
  pages={4620--4626},
  year={2025},
  publisher={ACS Publications}
}

@article{de2017non,
  title={Non-Poissonian photon statistics from macroscopic photon cutting materials},
  author={De Jong, Mathijs and Meijerink, Andries and Rabouw, Freddy T},
  journal={Nat. Commun},
  volume={8},
  number={1},
  pages={15537},
  year={2017},
  publisher={Nature Publishing Group UK London}
}

@article{milstein2018picosecond,
  title={Picosecond quantum cutting generates photoluminescence quantum yields over 100\% in ytterbium-doped {CsPbCl}\textsubscript{3} nanocrystals},
  author={Milstein, Tyler J and Kroupa, Daniel M and Gamelin, Daniel R},
  journal={Nano Lett.},
  volume={18},
  number={6},
  pages={3792--3799},
  year={2018},
  publisher={ACS Publications}
}

@article{kroupa2018quantum,
  title={Quantum-cutting ytterbium-doped {CsPb}({Cl}\textsubscript{1--x}{Br}\textsubscript{x})\textsubscript{3} perovskite thin films with photoluminescence quantum yields over 190\%},
  author={Kroupa, Daniel M and Roh, Joo Yeon and Milstein, Tyler J and Creutz, Sidney E and Gamelin, Daniel R},
  journal={ACS Energy Lett.},
  volume={3},
  number={10},
  pages={2390--2395},
  year={2018},
  publisher={ACS Publications}
}

@article{roh2023evolution,
  title={Evolution of {Yb}\textsuperscript{3+} Speciation in {Cl}\textsuperscript{--}/{Br}\textsuperscript{--}-and {Yb}\textsuperscript{3+}/{Gd}\textsuperscript{3+}-Alloyed Quantum-Cutting Lead-Halide Perovskite Nanocrystals},
  author={Roh, Joo Yeon D and Sommer, David E and Milstein, Tyler J and Dunham, Scott T and Gamelin, Daniel R},
  journal={Chem. Mater.},
  volume={35},
  number={19},
  pages={8057--8064},
  year={2023},
  publisher={ACS Publications}
}

@article{roh2020yb,
  title={{Yb}\textsuperscript{3+} speciation and energy-transfer dynamics in quantum-cutting {Yb}\textsuperscript{3+}-doped {CsPbCl}\textsubscript{3} perovskite nanocrystals and single crystals},
  author={Roh, Joo Yeon D and Smith, Matthew D and Crane, Matthew J and Biner, Daniel and Milstein, Tyler J and Kr{\"a}mer, Karl W and Gamelin, Daniel R},
  journal={Phys. Rev. Mater.},
  volume={4},
  number={10},
  pages={105405},
  year={2020},
  publisher={APS}
}

@article{erickson2019photoluminescence,
  title={Photoluminescence Saturation in Quantum-Cutting {Yb}\textsuperscript{3+}-Doped {CsPb}({Cl}\textsubscript{1--x}{Br}\textsubscript{x})\textsubscript{3} Perovskite Nanocrystals: Implications for Solar Downconversion},
  author={Erickson, Christian S and Crane, Matthew J and Milstein, Tyler J and Gamelin, Daniel R},
  journal={J. Phys. Chem. C},
  volume={123},
  number={19},
  pages={12474--12484},
  year={2019},
  publisher={ACS Publications}
}

@article{sommer2022defect,
  title={Defect formation in {Yb}-doped {CsPbCl}\textsubscript{3} from first principles with implications for quantum cutting},
  author={Sommer, David E and Gamelin, Daniel R and Dunham, Scott T},
  journal={Phys. Rev. Mater.},
  volume={6},
  number={2},
  pages={025404},
  year={2022},
  publisher={APS}
}

@article{li2019mechanism,
  title={Mechanism for the extremely efficient sensitization of {Yb}\textsuperscript{3+} luminescence in {CsPbCl}\textsubscript{3} nanocrystals},
  author={Li, Xiyu and Duan, Sai and Liu, Haichun and Chen, Guanying and Luo, Yi and {\AA}gren, Hans},
  journal={J. Phys. Chem. Lett.},
  volume={10},
  number={3},
  pages={487--492},
  year={2019},
  publisher={ACS Publications}
}

@article{xu2023atomic,
  title={Atomic-scale imaging of ytterbium ions in lead halide perovskites},
  author={Xu, Wen and Liu, Jiamu and Dong, Bin and Huang, Jindou and Shi, Honglong and Xue, Xiangxin and Liu, Mao},
  journal={Sci. Adv.},
  volume={9},
  number={35},
  pages={eadi7931},
  year={2023},
  publisher={American Association for the Advancement of Science}
}

@article{van2025cspbcl3,
  title={{CsPbCl}\textsubscript{3}:{Yb}\textsuperscript{3+} nanocrystals: Adverse effects of colloidally stable ytterbium-rich reaction by-products on luminescent down-conversion performance},
  author={Van de Voorde, Mathis and Hudry, Damien and Busko, Dmitry and Richards, Bryce S and Saive, Rebecca},
  journal={Opt. Mater.: X},
  volume={26},
  pages={100407},
  year={2025},
  publisher={Elsevier}
}

@article{zhao2020room,
  title={Room-temperature doping of ytterbium into efficient near-infrared emission {CsPbBr}\textsubscript{1.5}{Cl}\textsubscript{1.5} perovskite quantum dots},
  author={Zhao, Shuangyi and Zhang, Yubo and Zang, Zhigang},
  journal={Chem. Commun.},
  volume={56},
  number={43},
  pages={5811--5814},
  year={2020},
  publisher={Royal Society of Chemistry}
}

@article{ishii2020sensitized,
  title={Sensitized {Yb}\textsuperscript{3+} Luminescence in {CsPbCl}\textsubscript{3} Film for Highly Efficient Near-Infrared Light-Emitting Diodes},
  author={Ishii, Ayumi and Miyasaka, Tsutomu},
  journal={Adv. Sci.},
  volume={7},
  number={4},
  pages={1903142},
  year={2020},
  publisher={Wiley Online Library}
}

@article{karaveli2011spectral,
  title={Spectral tuning by selective enhancement of electric and magnetic dipole emission},
  author={Karaveli, Sinan and Zia, Rashid},
  journal={Phys. Rev. Lett.},
  volume={106},
  number={19},
  pages={193004},
  year={2011},
  publisher={APS}
}

@article{vergeer2005quantum,
  title={Quantum cutting by cooperative energy transfer in {Yb$_x$ Y$_{1-x}$ PO$_4$: Tb$^{3+}$}},
  author={Vergeer, Peter and Vlugt, TJH and Kox, MHF and Den Hertog, MI and Van der Eerden, JPJM and Meijerink, A},
  journal={Phys. Rev. B Condens. Matter},
  volume={71},
  number={1},
  pages={014119},
  year={2005},
  publisher={APS}
}

@article{senden2015photonic,
  title={Photonic effects on the radiative decay rate and luminescence quantum yield of doped nanocrystals},
  author={Senden, Tim and Rabouw, Freddy T and Meijerink, Andries},
  journal={ACS Nano},
  volume={9},
  number={2},
  pages={1801--1808},
  year={2015},
  publisher={ACS Publications}
}

@article{dodson2012magnetic,
  title={Magnetic dipole and electric quadrupole transitions in the trivalent lanthanide series: Calculated emission rates and oscillator strengths},
  author={Dodson, Christopher M and Zia, Rashid},
  journal={Phys. Rev. B},
  volume={86},
  number={12},
  pages={125102},
  year={2012},
  publisher={APS}
}

@article{rabouw2016europium,
  title={Europium-doped {NaYF$_4$} nanocrystals as probes for the electric and magnetic local density of optical states throughout the visible spectral range},
  author={Rabouw, Freddy T and Prins, P Tim and Norris, David J},
  journal={Nano Lett.},
  volume={16},
  number={11},
  pages={7254--7260},
  year={2016},
  publisher={ACS Publications}
}

@article{wang2017photonic,
  title={Photonic effects for magnetic dipole transitions},
  author={Wang, Zijun and Senden, Tim and Meijerink, Andries},
  journal={J. Phys. Chem. Lett.},
  volume={8},
  number={23},
  pages={5689--5694},
  year={2017},
  publisher={ACS Publications}
}

@article{zhang2024optimizing,
  title={Optimizing the Design of the Vapor-Deposited {CsPbCl$_3$}-Based Optoelectronic Devices via Simulations and Experiments},
  author={Zhang, Xuning and Chen, Linya and Liu, Xingyue and Liu, Zhiyong and Gu, Honggang and Bo, Sun and Liao, Guanglan},
  journal={Adv. Funct. Mater.},
  volume={34},
  number={7},
  pages={2310945},
  year={2024},
  publisher={Wiley Online Library}
}

@article{mangnus2021finite,
  title={Finite-size effects on energy transfer between dopants in nanocrystals},
  author={Mangnus, Mark JJ and Zom, Jeffrey and Welling, Tom AJ and Meijerink, Andries and Rabouw, Freddy T},
  journal={ACS Nanosci. Au},
  volume={2},
  number={2},
  pages={111--118},
  year={2021},
  publisher={ACS Publications}
}

@article{nedelcu2015fast,
  title={Fast anion-exchange in highly luminescent nanocrystals of cesium lead halide perovskites ({CsPbX$_3$}, {X=Cl, Br, I})},
  author={Nedelcu, Georgian and Protesescu, Loredana and Yakunin, Sergii and Bodnarchuk, Maryna I and Grotevent, Matthias J and Kovalenko, Maksym V},
  journal={Nano Lett.},
  volume={15},
  number={8},
  pages={5635--5640},
  year={2015},
  publisher={ACS Publications}
}

@article{guhrenz2016solid,
  title={Solid-state anion exchange reactions for color tuning of {CsPbX$_3$} perovskite nanocrystals},
  author={Guhrenz, Chris and Benad, Albrecht and Ziegler, Christoph and Haubold, Danny and Gaponik, Nikolai and Eychmuller, Alexander},
  journal={Chem. Mater.},
  volume={28},
  number={24},
  pages={9033--9040},
  year={2016},
  publisher={ACS Publications}
}

@article{roh2023negative,
  title={Negative Thermal Quenching in Quantum-Cutting {Yb}\textsuperscript{3+}-Doped {CsPb}({Cl}\textsubscript{1--x}{Br}\textsubscript{x})\textsubscript{3} Perovskite Nanocrystals},
  author={Roh, Joo Yeon D and Milstein, Tyler J and Gamelin, Daniel R},
  journal={ACS nano},
  volume={17},
  number={17},
  pages={17190--17198},
  year={2023},
  publisher={ACS Publications}
}

@article{tepliakov2024trap,
  title={Trap-Mediated Sensitization Governs Near-Infrared Emission from Yb$^{3+}$-Doped Mixed-Halide {CsPbCl$_x$Br{$_{3-x}$}} Perovskite Nanocrystals},
  author={Tepliakov, Nikita V and Sokolova, Anastasiia V and Tatarinov, Danila A and Zhang, Xiaoyu and Zheng, Weitao and Litvin, Aleksandr P and Rogach, Andrey L},
  journal={Nano Lett.},
  volume={24},
  number={11},
  pages={3347--3354},
  year={2024},
  publisher={ACS Publications}
}

@article{zhou2017cerium,
  title={Cerium and ytterbium codoped halide perovskite quantum dots: a novel and efficient downconverter for improving the performance of silicon solar cells},
  author={Zhou, Donglei and Liu, Dali and Pan, Gencai and Chen, Xu and Li, Dongyu and Xu, Wen and Bai, Xue and Song, Hongwei},
  journal={Adv. Mater.},
  volume={29},
  number={42},
  pages={1704149},
  year={2017},
  publisher={Wiley Online Library}
}

@article{pan2017doping,
  title={Doping lanthanide into perovskite nanocrystals: highly improved and expanded optical properties},
  author={Pan, Gencai and Bai, Xue and Yang, Dongwen and Chen, Xu and Jing, Pengtao and Qu, Songnan and Zhang, Lijun and Zhou, Donglei and Zhu, Jinyang and Xu, Wen and others},
  journal={Nano Lett.},
  volume={17},
  number={12},
  pages={8005--8011},
  year={2017},
  publisher={ACS Publications}
}

@article{zhang2018yb,
  title={{Yb}$^{3+}$ and {Yb}$^{3+}$/{Er}$^{3+}$ doping for near-infrared emission and improved stability of {CsPbCl$_3$} nanocrystals},
  author={Zhang, Xiangtong and Zhang, Yu and Zhang, Xiaoyu and Yin, Wenxu and Wang, Yu and Wang, Hua and Lu, Min and Li, Zhiyang and Gu, Zhiyong and Yu, William W},
  journal={J. Mater. Chem. C},
  volume={6},
  number={37},
  pages={10101--10105},
  year={2018},
  publisher={Royal Society of Chemistry}
}

@article{luo2018quantum,
  title={Quantum-cutting luminescent solar concentrators using ytterbium-doped perovskite nanocrystals},
  author={Luo, Xiao and Ding, Tao and Liu, Xue and Liu, Yuan and Wu, Kaifeng},
  journal={Nano Lett.},
  volume={19},
  number={1},
  pages={338--341},
  year={2018},
  publisher={ACS Publications}
}

@article{zhou2019impact,
  title={Impact of host composition, codoping, or tridoping on quantum-cutting emission of ytterbium in halide perovskite quantum dots and solar cell applications},
  author={Zhou, Donglei and Sun, Rui and Xu, Wen and Ding, Nan and Li, Dongyu and Chen, Xu and Pan, Gencai and Bai, Xue and Song, Hongwei},
  journal={Nano Lett.},
  volume={19},
  number={10},
  pages={6904--6913},
  year={2019},
  publisher={ACS Publications}
}

@article{ye2023980,
  title={980 nm Near-Infrared Light-Emitting Diode Using All-Inorganic Perovskite Nanocrystals Doped with Ytterbium Ions},
  author={Ye, Zhenglan and Liu, Taoran and Chen, Dan and Yang, Yazhou and Li, Jiayi and Pang, Yaqing and Liu, Xiangquan and Zuo, Yuhua and Zheng, Jun and Liu, Zhi and others},
  journal={Tsinghua Sci. Technol.},
  volume={29},
  number={1},
  pages={207--215},
  year={2023},
  publisher={TUP}
}

@article{cleveland2023physical,
  title={Physical vapor deposition of {Yb}-doped {CsPbCl$_3$} thin films for quantum cutting},
  author={Cleveland, Iver J and Tran, Minh N and Kabra, Suryansh and Sandrakumar, Kajini and Kannan, Haripriya and Sahu, Ayaskanta and Aydil, Eray S},
  journal={Phys. Rev. Mater.},
  volume={7},
  number={6},
  pages={065404},
  year={2023},
  publisher={APS}
}

@article{stefanski2021optical,
  title={Optical characterization of {Yb$^{3+}$}: {CsPbCl$_3$} perovskite powder},
  author={Stefanski, M and Ptak, M and Sieradzki, A and Strek, W},
  journal={Chem. Eng. J.},
  volume={408},
  pages={127347},
  year={2021},
  publisher={Elsevier}
}

@article{demkiv2024effect,
  title={Effect of {Yb} doping on the optical and photoelectric properties of {CsPbCl$_3$} single crystals},
  author={Demkiv, TM and Chornodolskyy, YaM and Muzyka, TM and Malynych, SZ and Serkiz, R Ya and Pushak, AS and Kotlov, A and Gamernyk, RV},
  journal={Opt. Mater.: X},
  volume={22},
  pages={100303},
  year={2024},
  publisher={Elsevier}
}

@article{rubio2026defect,
  title={Defect Tolerant Quantum Cutting in Mechanosynthesized Ytterbium-Doped Cesium Lead Chloride Perovskites},
  author={Rubio, Thiago I and Avalos, Claudia E},
  journal={Chem. Mater.},
  year={2026},
  publisher={ACS Publications}
}

@article{mukherjee2026mechanistic,
  title={Mechanistic Insights into Quantum-Cutting in {Yb}$^{3+}$-Doped {CsPbCl$_3$} Nanocrystals},
  author={Mukherjee, Poulomi and Sarma, DD},
  journal={Small},
  volume={22},
  number={27},
  pages={e14834},
  year={2026},
  publisher={Wiley Online Library}
}

@article{milstein2019anion,
  title={Anion Exchange and the Quantum-Cutting Energy Threshold in Ytterbium-Doped {CsPb (Cl$_{1-x}$Br$_x$)$_3$} Perovskite Nanocrystals},
  author={Milstein, Tyler J and Kluherz, Kyle T and Kroupa, Daniel M and Erickson, Christian S and De Yoreo, James J and Gamelin, Daniel R},
  journal={Nano Lett.},
  volume={19},
  number={3},
  pages={1931--1937},
  year={2019},
  publisher={ACS Publications}
}

\end{document}